\documentclass[12pt,preprint]{aastex}

\begin{document}

\newcommand{\eg}{e.g.,\,}
\newcommand{\ie}{i.e.,\,}
\newcommand{\etal}{et al.}
\newcommand{\be}{\begin{equation}}
\newcommand{\ee}{\end{equation}}
\newcommand{\bea}{\begin{eqnarray}}
\newcommand{\eea}{\end{eqnarray}}
\newcommand{\checkit}{\fbox{{\tiny$\surd$}}}

\title{Chang-Refsdal Microlensed Type Ia Supernova Light Curves}
\author{
H. Bagherpour\footnote{hamed@nhn.ou.edu} , 
B. Chen 
\footnote{chen@nhn.ou.edu} ,
R. Kantowski
\footnote{kantowski@nhn.ou.edu} ,
D. Branch 
\footnote{branch@nhn.ou.edu}
}

\affil{University of Oklahoma, Homer. L. Dodge Department of Physics
 and Astronomy,\\ Norman, OK 73019, USA }

\author{D. Richardson 
\footnote{richardson@denison.edu}
}

\affil{Denison University, Department of Physics and Astronomy, \\
 Granville, OH 43023, USA }

\begin{abstract}

A description of the light curves of microlensed Type Ia supernovae
(SNe~Ia) as extended and expanding sources is presented.  We give
examples of what microlensing by stellar-mass Chang-Refsdal lenses can
do to a small percentage of supernova light curves.  We find that in
addition to overall brightening, significant changes in light--curve
shapes can also occur.  Peaks can be distorted, plateaus can appear,
and even secondary peaks can be formed.  The effects of both the
relative motion of the lens and the supernova and the expansion of the
supernova are given and compared.  The effects of relative motion are
more pronounced when a distant supernova ($z_s \sim 1$) impacts well
within the Einstein ring of a nearby microlens ($z_d \sim 0.05$) and
are less important for more distant deflectors.  We also find that the
increase in shear that comes with increased deflector distance tends
to reduce the time variability of microlensing.  We briefly discuss
the probability of observing these effects.

\end{abstract}


\section{Introduction }

Out of a number of distance indicators, Type~Ia supernovae (SNe~Ia)
have emerged as the most promising standard candles. Due to their
significant intrinsic brightness and relative ubiquity they can be
observed in the local and distant universe. Several teams including
the High-z Supernova Search \citep{Schmidt98} and the Supernova
Cosmology Project \citep{Perlmutter99} have been searching for SNe~Ia
at higher redshifts since the early 1990's. Light emitted from these
`standard candles' is subject to lensing by intervening objects while
traversing the large distances involved \citep{KVB95}; the further the
light source, the higher its chance of being significantly lensed.  In
fact, for cosmologically distant sources, the probability is high that
a distant point source will be `imaged' \citep{Press73,BK76,Wyithe02},
particularly by stellar objects (microlensing).  While the
systematic errors introduced by K-correction, selection effects, and
possible evolution can be removed, lensing might ultimately limit the
accuracy of luminosity distance measurements \citep{Perlmutter03}.
Only a large sample of SNe~Ia at each redshift can be used to
characterize the lensing distribution and to correct for the effect of
weak lensing. Proposed searches, e.g., \cite{Tyson05, Corasaniti06} are expected to yield hundreds of thousands of SNe~Ia out
to $z \sim 1$, some of which should exhibit lensing effects such as
described here.

Properties of microlensed supernovae have been studied previously.
\citet{Schneider87} presented the time-dependent amplification of
supernovae caused by their expansion and showed that the related
polarization is not likely to exceed 1\% and hence not possible to
detect among cosmologically distant supernovae at this time.
\citet{Linder88} studied amplification of supernovae and developed
approximate formulae for the amplification probability
distribution. \citet{Rauch91} studied microlensing of SNe Ia by
compact objects and calculated the resulting amplification probability
distributions using Monte Carlo simulations.  Assuming
point-deflectors with shear \citet{Kolatt98} estimated the
microlensing rate by MACHOs in nearby clusters. The supernovae were
treated as point sources and consequently the microlensed light curves
were expected to contain amplification peaks similar to caustic
crossing events seen in the OGLE data \citep{Udalski03}.  In this
paper, we demonstrate in detail how microlensing by a single
stellar deflector in an external shear field can affect light
curves of cosmologically distant SNe Ia. We use the lens model of
\citet{Chang79} to calculate amplifications \citep{Schneider92}.  We
ignore any lensing amplification caused by the SN's hosting galaxy and
concentrate on the time-dependent effects caused by a single moving
stellar deflector.  We model the SNe Ia as expanding light
sources with limb-darkening.
  
In the next section, we present our model for SNe~Ia as microlensed
sources.  Section 3 is devoted to a brief discussion of microlensing
of finite-size sources, and in $\S$ 4 we present the results of our
calculations.  Throughout, we assume a flat
Friedmann-Lama\^{\i}tre-Robertson-Walker cosmological model with
$\Omega _{m}=$ 0.3, $\Omega _{\Lambda }=$ 0.7, and h$_{100}=$ 0.67 to
calculate the distances to the source and deflector.

\section{Type Ia Supernovae}
\subsection{The Light Curve}

To model the intrinsic SN~Ia light curve we used a combination of two
analytical models; that of \citet{Arnett82} for the peak of the light
curve (the photospheric phase) and that of \citet{Jeffery99} for the
tail (the nebular phase).  The details of this model are in
\citet{Richardson05}. The model parameters have been fixed for a
typical SN~Ia: the kinetic energy is $10^{51}$~erg), the total ejected
mass is 1.4~M$_{\odot }$, and the $^{56}$Ni mass has been set to 0.6
M$_{\odot }$. These values produce a peak absolute magnitude of $M_B =
-19.5$ and a light-curve shape that conforms to the characteristic
light--curve of normal SNe~Ia, see the solid curves in the first rows of Figures 3-5.

\subsection{The Radius}

Because the effective size of a SN~Ia varies with time and is on
occasion of the same dimension as the lens caustics we need an
expression for its time-dependent radius.  For the photospheric phase
($ t < 150$ days) we obtain the radius by multiplying the velocity at
the photosphere of the homologously expanding ejecta, as determined
empirically by \citet{Branch05}, by the time since explosion.  For the
nebular phase, we assume that the effective radius of the iron--group
core expands at a constant velocity of 6000 km s$^{-1}$.  A good fit
for the velocity is then an exponential:
\begin{equation} 
v(t)=9.1\, \textrm{e}^{\left(-\textrm{t}/\textrm{36.5}\, \textrm{days}\right)}+6.0 \: ,
\end{equation}  
in units of $10^{3}\ \textrm{km s}^{-1}$, from which the photospheric radius in AU is given by:
\begin{equation} 
r_{SN}(t)=v(t)\,t=\left[ 5.3\,\textrm{e}^{\left(-\textrm{t}/\textrm{36.5}\, \textrm{days}\right)}+3.5\right]\,t \: .
\label{rpht}
\end{equation}  
Figure 1 shows the expansion velocity as well as the radius as a function of time.

\subsection{Limb-darkening}

For limb darkening in the nebular phase we assume uniform
emissivity per unit volume (out to 6000~km~s$^{-1}$) and find
\begin{equation}
I(r) = I_{o} \sqrt {1-\left( \frac{r}{r_{SN}} \right)^{2}}\: ,
\label{intensity}
\end{equation}
where $I_{o}$ is the intensity at the center and $r_{SN}$ is the
supernova's radius. The assumption of uniform emissivity per unit volume is a
reasonable assumption during the nebular phase; to see whether the
same expression can be used for the photospheric phase we compare it
in Figure~2 to limb-darkening curves in the U and B bands calculated
(courtesy of E.~Lentz) for the W7 model \citep{Nomoto84}.  Considering
that the detailed calculations are model dependent, the use of the
simple nebular-phase expression for the photospheric phase is
reasonable.

\section{Microlensing of Extended Sources }

\subsection{Basics}

The linearized Einstein theory for a static gravitational field gives
a bending angle for light rays passing through a weak gravitational
field of 
\begin{equation}
\mbox{\boldmath{$\alpha$}}=-\frac{2}{c}\int _{-\infty }^{+\infty }
\mbox{\boldmath{$\nabla$}}\phi \, dt\: ,\label{eq:1}
\end{equation}
where $\phi$ is the Newtonian gravitational potential satisfying
the boundary conditions $\phi \rightarrow 0$ at infinity, and where
the integral is performed along the light path in the absence of the
gravitational field. \citet{BKN73}, and
\citet{BK74,BK76} used the 2-component nature of
\mbox{\boldmath{$\alpha$}} to replace it with the complex scattering function $I(z)$,
where $z=x+iy$ is the complex equivalent of the 2-d vector $\mathbf{r}=x\hat{\imath }+y\hat{\jmath }$.
Using $I(z)$, the relation between the source position, $z$, and
image position, $z_{o}$, when both are projected onto the plane of
the deflector is 
\begin{equation}
z=z_{o}-\frac{4GD}{c^{2}}I^{*}(z_{o})\: ,
\label{z=z0}
\end{equation}
where * means complex conjugate.
The scaled (effective) distance $D$ is defined as $D=D_{ds}D_{d}/D_{s}$, where
the deflector-source distance, $D_{ds}$, the observer-deflector distance,
$D_{d}$, and the observer-source distance, $D_{s}$, are all the
same type distances, e.g., apparent size distances.

The effect of gravitational lensing on the apparent brightness of
a distant source can be computed in various ways. For extended sources
it is often easiest to employ the fact that the apparent brightness
is proportional to the image's apparent area, i.e., the brightness
of a small source is amplified by a factor 
\begin{equation}
A=\frac{{\cal A}_{o}}{{\cal A}}\: ,
\label{A}
\end{equation}
where ${\cal A}_{o}$ is the area of the image and ${\cal A}$ is the area of the
source, both projected on the deflector plane.
\subsubsection{The Schwarzschild Lens}
In the case of an isolated point deflector,
the scattering function takes a very simple form: 
\begin{equation}
I(z_{o})=\frac{m_{d}}{z_{o}}\: ,\label{eq:4}
\end{equation}
and the equation (\ref{z=z0}) reduces to 
\begin{equation}
z=z_{o}-\frac{r_{E}^{2}}{z_{o}^{*}}\: ,
\label{SS-lens}
\end{equation}
where $r_{E}\equiv \sqrt{4Gm_{d}D\, c^{-2}}=\sqrt{2r_{S} D}$ is the
Einstein ring radius, and $r_{S}$ is the deflector's Schwarzschild
radius. This equation has two separate solutions (images) for any
source position $r=|z|$, 
\begin{equation}
r_{\pm}=\left|z_{\pm}\right|=\frac{1}{2}\left(\sqrt{r^{2}+4r_{E}^{2}} \pm r\right)\: .
\label{SS-roots}
\end{equation}
Both images are in line with source and deflector. The `primary'
image, $r_{+}$, lies on the same side of the deflector while the
`secondary' image, $r_{-}$, is on the other side. In
the case of microlensing, the angular separation of the two images
 is of the order of micro
arcseconds and consequently seen as a single object. 
The Einstein ring occurs when source and deflector are aligned with
the observer $(r=\left|z\right|=0)$ for which $r_{+}=r_{-}=r_{E}$, and due 
to the symmetry of the lensing configuration the image is
actually a ring. 

 For a point mass deflector
and a small point-like source the combined amplification of the unresolved primary and
secondary images as a function of the image positions $r_\pm$ or the source position $r$ is 
\begin{equation}
A \equiv A_{+}+A_{-}=\left|\frac{1}{1-r_E^4/r_+^4}\right|+\left|\frac{1}{1-r_E^4/r_-^4}\right|=\frac{r^{2}+2r_{E}^{2}}{r\sqrt{r^{2}+4r_{E}^{2}}}\: .
\label{Ap+As}
\end{equation}

Measuring this amplification from a single observation is not possible
since it is practically impossible to figure out the original source
flux. However, if the luminosity of the source varies with time in
a predictable way as with SNe or if the source is of constant brightness
and the lens is moving with respect to the line of sight to the source
[as with observation of bulge stars; see, for instance, \citet{Sumi04}],
the amplification will change with time in a predictable way
and it is possible to determine the amplification.
\subsubsection{The Chang-Refsdal Lens}

If a point mass lens is not isolated but instead lives in a gravity field 
which varies slowly on the length scale of $r_E$ then the lens equation (\ref{SS-lens})
changes to 
\be
z=z_o-\frac{r_E^2}{z_o^*}-\gamma z_o^*-\kappa z_o ,
\label{CR-lens}
\ee
where $\gamma$ is the complex shear and $\kappa$ the convergence of the local gravity field \citep{Witt90,Mao91}.
The convergence is proportional to the transparent surface mass density in the immediate 
neighborhood of the microlensing star and will be assumed to vanish in what we do here.
The effects of a small convergence are easily accounted for and are not qualitatively different from $\kappa=0$.
Shear is introduced by the macro-lens structure of the galaxy as a whole as well as the nearby neighbors 
of the microlensing star, see \cite{Nityananda84,Totani03}. It can introduce significant qualitative differences, \eg if the source 
is within a 2-d domain bounded by the diamond shaped caustic ($|\gamma|<1$, see rows 4 of Figures 3-5) there are 2 extra 
images not present with the isolated star. 
The two extra images exist because Eq.\,(\ref{CR-lens}) has 4 solutions $z_o=z_1,z_2,z_3,z_4$ when 
the source $z$ is inside the closed caustic curve and only 2 solutions when it is not. Expressions for 
the 4 (or 2) image positions equivalent to Eq.\,(\ref{SS-roots}) exist (they are given by the 4 (or 2) real solutions of a quartic equation in $r_o^2\equiv z_oz_o^*$) but are more complicated than Eq. (\ref{SS-roots}). 
The expression for the amplification of each image is relatively simple when written as a function of the image's position $z_o$ 
\be
A^{-1}=\left| (1-\kappa)^2-r_E^4/r_o^4-\gamma\gamma^*+\gamma(r_E/z_o)^2+\gamma^*(r_E/z_o^*)^2\right|.
\label{CR-amp}
\ee
However, when written as a function of the source position $z$, $A^{-1}$ is quite complicated and the equivalent 
of Eq.\,(\ref{Ap+As}) is even worse. The critical curve (somewhat elliptically shaped when $\kappa+|\gamma|<1$) for the Chang-Refsdal lens (equivalent to the circularly shaped Einstein ring for the Schwarzschild lens) is drawn by putting $A^{-1}\rightarrow 0$ in Eq.\,(\ref{CR-amp})
and from that curve the diamond shaped caustic is drawn using Eq.\,(\ref{CR-lens}), see row 4 of Figures 3-6. The minor-major axes 
for the critical curves and caustics are (in units of $r_E$) respectively 
$1/\sqrt{1-\kappa\pm|\gamma|}$
and $2|\gamma|/\sqrt{1-\kappa\pm|\gamma|}$ and are oriented along the x-y axes rotated counterclockwise through an angle of half the phase of $\gamma$.

\subsection{Amplification of an Extended Source}

If the source is not small, \eg for the Schwarzschild lens, if $r/r_{E}$ varies significantly
across the source, differential amplification must be taken into account.
In general to obtain the total flux received
from an extended source, an integral of intensity $I$ across the source
may be required: 
\begin{equation}
A=\frac{\int _{images} I\, d{\cal A}_{o}}{\int _{source} I\, d{\cal A}}\: .\label{eq:10}
\end{equation}
If the surface brightness is constant across the source or if there
is no differential amplification, the net amplification is simply
given by equation (\ref{A})  where ${\cal A}_{o}$ can be any of the images areas
or in fact the total image area (giving the total amplification).

\subsubsection{The Schwarzschild Lens}
The amplification of a  disk source with constant surface brightness, lensed by an isolated point 
mass can be given analytically.
If the circular extended source has 
a projected radius of $a$ and is at a distance $l$ from the center of a
spherically symmetric deflector ($a$ and $l$ are measured in the
deflector plane), the total area of the combined and unresolvable Schwarzschild images
is 
\begin{equation}
{\cal A}^{(total)}=\int _{-\frac{\pi }{2}}^{\frac{\pi }{2}}a\left(a+l\, \sin \varphi \right)
\sqrt{1+\frac{4r_{E}^{2}}{l^{2}+a^{2}+2al\, \sin \varphi }}d\varphi \: 
\end{equation}
After a rather long calculation, the total amplification of a uniform disc is found
to be 
\begin{equation}
A_{disc}[a,\, l,\, r_{E}]=\eta \left\{ \mu _{1}K\left(k\right)+\mu _{2}E\left(k\right)+
\mu _{3}\Pi \left(n,\, k\right)\right\} \: ,
\label{Aellip}
\end{equation}
\citep{Witt94,Mao98} where $K$, $E$, and $\Pi $ are respectively
the first, second, and third complete elliptic integral with 
\[
k=\frac{16alr_{E}^{2}}{\left(l+a\right)^{2}\left(\left(l-a\right)^{2}+4r_{E}^{2}\right)}\: ,
\]
\[
n=\frac{4al}{\left(a+l\right)^{2}}\: ,
\] 
and constants 
\[
\eta =\frac{1}{2\pi a^{2}\sqrt{\left(l-a\right)^{2}+4r_{E}^{2}}}\: ,
\]
\[
\mu _{1}=\left(l-a\right)\left(a^{2}-l^{2}-8r_{E}^{2}\right)\: ,
\]
\[
\mu _{2}=\left(l+a\right)\left(\left(l-a\right)^{2}+4r_{E}^{2}\right)\: ,
\]
\[
\mu _{3}=\frac{4\left(l-a\right)^{2}\left(a^{2}+r_{E}^{2}\right)}{l+a}\: . 
\]
The above applies to the so called Schwarzschild lens where shear, $\gamma$, is negligible 
and the deflector's size is sufficiently small.

The amplification of a thin ring of radius $a$ can be 
computed using Eq. (10) as
\begin{equation}
A_{ring}\left( a, l, r_{E} \right)=\frac{1}{2\pi a}\frac{d}{da}\left[ \pi a^{2}  
A_{disc}\left( a, l, r_{E} \right) \right] \: ,
\end{equation}
and the net amplification for a limb-darkened source where photospheric radius is $r_{ph}$ as
\begin{equation}
A_{I}\left( r_{ph}, l, r_{E} \right)=
\frac{-\int_{0}^{r_{ph}}\frac{d}{da}\left[ I(a) \right] \pi a^{2} A_{disc}\left( a, l, r_{E} \right) da}
{\int_{0}^{r_{ph}} I(a) 2\pi a da}\: .
\end{equation}
For the intensity profile $I(r)$ given in Eq.\,(\ref{intensity}), $A_{I}$ simplifies to:
\begin{equation}
A_{I}\left( r_{ph}, l, r_{E} \right)=\frac{3}{2}\int_{0}^{\pi/2}\sin^{3}\varphi\, 
A_{disc}[r_{ph}\sin \varphi,\, l,\, r_{E}]\, d\varphi\, ,
\label{AI}
\end{equation}
where $A_{disc}$ is the function introduced in Eq. (\ref{Aellip}).

\subsubsection{The Chang-Refsdal Lens}

For the Chang-Refsdal lens we resort to a Monte Carlo calculation, randomly covering the extended source 
with points and summing over each image amplification using Eq.\,(\ref{CR-amp}) to compute the net amplification $A$.
Even though Eq.\,(\ref{AI}) is valid for the Chang-Refsdal lens 
the absence of a known analytic expression for $A_{disc}$, good for arbitrary source sizes, prohibits its use. \citet{Schneider87} have given an analytic expression for the amplification of a small uniform disc near a long critical line,
however, for moving and expanding SN sources this condition is short lived.
The Monte Carlo calculation is easily adapted to account for an expanding and limb-darkened source
and the $\gamma\rightarrow 0$ results agree with results from the above Schwarzschild lens.
In most cases we were able to obtain accurate values for the amplification by using less than 100,000 points;
however, we did have to eliminate noise from the Poisson statistics on occasion.

\subsection{Probability }

The relevant quantity in seeing a microlensing event is the optical
depth $\tau $. It is defined as the probability that a point source
(or equivalently the center of an extended source) falls inside the Einstein ring
of some deflector. The brightness of a point source within
$r_{E}$ of a Schwarzschild lens is amplified by a factor of at least 1.34. For randomly located point deflectors
the optical depth depends on the mass density of the deflectors 
and not on their number density \citep{Press73}.
Typical values of the optical depths for microlensing of nearby stellar
sources are remarkably small. For instance, \citet{Sumi03}
gives an optical depth $\tau =2.59_{-0.64}^{+0.84}\times 10^{-6}$
toward the Galactic Bulge (GB) in Baade's window for events with time
scales between 0.3 and 200 days. Because of the small value of $\tau $,
millions of stars need to be monitored when searching for microlensing
in areas such as the GB or the Large and Small Magellanic Clouds. The value of $\tau $ is much higher
for cosmologically distant sources. 

If we look at bulge-bulge lensing , \citet{Han03} give a model 
for the bulge from which a value of 
$\tau=0.98\times 10^{-6}$ is computed. They additionally compute  
an effective column density for deflectors $\Sigma_*=$ 2086 M$_{\sun}$pc$^{-2}$ and a 
characteristic source-lens separation $\overline D = 782$\,pc defined by 
\begin{equation}
\tau=\frac{4\pi G}{c^2}\Sigma_*\overline D.
\end{equation}
If we now look at the bulge of a similar galaxy at $z= 0.05$ we
expect a similar $\Sigma_*$ but $\overline D$ becomes the distance to
the deflector $D_d$ and hence increased by a factor $\approx 2.7\times
10^5$. This would bring the optical depth up to ~27\% when looking
through such a galaxy.  At this distance the size of the bulge is
$\sim 1$ arcsec and clearly resolvable. The downside is one of
alignment. What is the chance of a galaxy hosting a SN being
appropriately aligned with a foreground galaxy?  The best place to see
this effect seems to be the foregrounds of dense clusters. Because the
SN~Ia rate in the typical galaxy at $z = 1$ is about one per hundred
years, some $370(1+z)$ alignments would have to be followed for a year
to see one event.

Besides the work of \citet{Rauch91} and \citet{Kolatt98} mentioned earlier, 
others have made detailed estimates of micro-lensing probabilities.
Assuming that the ordinary stellar
populations of galaxies are the dominant causes of microlensing events,
\citet{Wyithe02} concluded that in a flat universe, at least
1\% of high-redshift sources ($z_{s}\geqslant 1$) are microlensed by stars
at any given time. \citet{Zakharov04} estimated that the optical
depth for microlensing caused by deflectors both localized in galaxies
and distributed uniformly, might reach 10\%  for sources at $z_{s}\sim 2$.
Assuming low mass Chang-Refsdal deflectors
\citet{Kolatt98} estimated that the microlensing rate of SNe by MACHOs
in nearby clusters ($z\le 0.05$) would be $\sim 0.02(f/0.01)$ per year
where $f$ is the fraction of the cluster mass in MACHOs ($10^{-7}\le
M_{macho}/M_\sun<10^{-4}$).
 
As mentioned above, the amplification of a point source falling inside
the Einstein ring $r_{E}$ of a Schwarzschild lens is larger than
1.34. The probability of a larger amplification is proportionally
smaller, \eg the probability of having an amplification larger than
$A$ for a given lensing configuration is
\begin{equation}
p(A)=u_{A}^{2}\, \tau(z_{s}) \: ,
\label{p(A)}
\end{equation} 
where $u_{A}\equiv b_{A}/r_{E}$ is the normalized impact parameter,
that results in amplification $A$ \citep{Paczynski86a,Paczynski86b} and $\tau(z_{s})$
is the optical depth for a source at redshift $z_{s}$. For a point
source, the result is
\begin{equation}
u_A=\sqrt{2\left(\frac{A}{\sqrt{A^{2}-1}}-1\right)} \: .
\label{u_A}
\end{equation}
The standard optical depth $\tau $ significantly
overestimates the probability that the interesting cases 
discussed in $\S$ 4 will occur. Using Eq.\,(\ref{p(A)}) or Eq.\,(\ref{u_A})
 we can try to correct for the overestimate. For these
lensing configurations, the normalized impact parameter $u_{A}$ does not
exceed 0.1 ($A\approx 9$ for the point source) which
gives a maximum probability 
of $\sim 10^{-4}$ for $z_{s}\geqslant 1$ ($\tau =0.01$) if \citet{Wyithe02} are correct, 
and $\sim 10^{-3}$
for $z_{s}=2$ ($\tau =0.1$) if \citet{Zakharov04} are correct.
These numbers above are somewhat higher than \citet{Linder88}
were predicting for Type I SN but not for Type II.
Probabilities likes these, together with time scales of some cases studied here, imply that
such effects may not be observed unless a large number of cosmologically distant
(around $10^{4}$ for $z_{s}\geqslant 1$) SNe~Ia are followed 
for a period of up to 2 years. 

Numerous probability estimates for weak (single-image) and strong (multi-image) macro-lensing of supernova have been made for 
possible deep searches \citep{Wang00,Holz01,Goobar02,Amanullah03}.
\citet{Oguri03} sumarizes that 0.05-0.1 \%  of the SN observed by SNAP 
at $z\sim 1$ would be macro-lensed. Such estimates can be used to indicate micro-lensing probabilities because a significant number of the deflectors for such  observations
would be at $z_d\approx 0.35$ and hence at high optical depth for micro-lensing. 
Large amplifications can also cause a bias in favor of observing supernovae, allowing
one to observe more distant objects, and as a result, to increase
the depth of any supernova survey.  \citet{Gunnar03} estimate that huge increases (\eg 100\%) in observable SN to 27th magnitude could occur if one simply searches in the direction of large clusters. 

For the high amplification cases seen in the next section where the micro-lensed light-curves have a second peak,
a separate probability estimate, consistent with the above, is given in the Appendix.

\section{Microlensed Light Curves }

In this section we use the SN Ia model  of $\S$ 2 and the
microlensing theory of $\S$ 3 to predict the shape of lensed light
curves of SNe~Ia for moving lenses. 
We have calculated absolute magnitudes
of the lensed light curves in the V-band. We concentrate on sources at
redshifts $z_{s}=1.0$ which are lensed by Chang-Refsdal deflectors at redshifts
$z_{d}=0.05, 0.10,$ and 0.35. We use $z_d=0.35$ because most lensing is expected to occur at this redshift, \ie $D=D_d D_{ds}/D_s$ is $\sim$ maximum.
To calculate
the amplification as a function of time we need
to know the distance $l(t)$ between the supernova's center and the deflector,
projected on the plane of the deflector, 
\begin{equation}
l(t)=\sqrt{l_{o}^{2}+vt\left(vt\mp 2\sqrt{l_{o}^{2}-b^{2}}\right)}\: ,\label{eq:24}
\label{l(t)}
\end{equation}    
where the minus sign is used when the supernova source explodes ($t=0$
and $l=l_{o}$) before getting to the point of closest approach, $b$,
and the plus sign when it explodes after. 
We also need the time dependent supernova's radius Eq.(\ref{rpht}),
and the limb-darkening expression Eq.(\ref{intensity}), both projected onto the deflector's plane.
The time in Eq.\,(\ref{l(t)}) is deflector time $t_d$ whereas the time in Eq.(\ref{rpht})
is source time $t_s=(1+z_d)/(1+z_s)\, t_d$ and both must be appropriately redshifted to be displayed in observer time as in Figures 3-5.
When the shear vanishes we used the Schwarzschild lens result Eq.(\ref{AI}), and
when $\gamma \ne 0$ we resorted to a Monte-Carlo procedure as indicated in $\S 3.2.2$. Because $\gamma$
is orientation dependent, microlensing depends on both its magnitude and direction, 
see Eq.\,(\ref{CR-lens}). 
However,
the resulting amplification Eq.\,(\ref{CR-amp}) is sensitive to orientation mainly when the projected 
photosphere is smaller than the caustic's dimension. We are able to give a representative sample of 
SN lensing by aligning the caustic with the $y$-axis and having the deflector 
move parallel to the x-axis.

We plot light curves for various values of the parameters $m_{d}$, $b$,
$l_{o}$, $v$ (the relative speed of source and deflector projected
on the deflector's plane) and $|\gamma|$ (the magnitude of the shear caused by the deflector's neighbors 
and/or the galaxy as a whole). 
Our sample deflector masses are $10^{-3}$ M$_{\odot }$,
1 M$_{\odot }$, and 10 M$_{\odot }$.  We fix the origin on the deflector and take the 
source to move in the deflector plane for 1,000 days (observer time) with 
three relative projected speeds
of the source $v$, 0.1\, AU/day, 0.5 AU/day, and 1.0 AU/day = 1,730 km/sec.

Figures 3 through 5 show the light curves ($1^{st}$ row of panels) and
the amplification curves ($2^{nd}$ row of panels) for lensed $z_s=1$
supernovae with a total of six different configurations of moving deflectors at
$z_d=0.05$.  The abscissas for the first 3 rows are in days since
explosion.  Each figure is for a fixed mass and contains two lensing
configurations (left and right columns).  Each panel of the first 2
rows show curves corresponding to 3 values of shear ($\gamma=0$ \ie
the Schwarzschild lens, and two values of $\gamma \ne 0$ \ie the
Chang-Refsdal lens). These interesting cases have been selected from a
number of configurations. Each panel in the $3^{rd}$ row contains six
curves of magnitude differences and show the relative importance of
the deflector-lens motion versus photospheric expansion as a function
of redshift. We have plotted the lensed magnitudes with relative
transverse deflector motion minus lensed magnitudes without motion for
two lenses at three redshifts, $z_d=0.05, 0.10,$ and 0.35.  Because
the local shear of a given macro-lens varies with deflector distance
$z_d$, we have appropriately varied $\gamma$ in the $3^{rd}$ row. The
two starting values of $\gamma$ are zero and one of the two values in
the first 2 rows (either $\gamma=$ 0.20 or 0.15) where $z_d=0.05$. At
$z_d=0.10$ and 0.35, $\gamma$ is scaled by 1.76 and 3.15 respectively.
In the $4^{th}$ row of 4 panels we show the deflector plane position
and size (in units of $r_E$) of the moving supernova's photosphere
relative to the caustics at two or three different times: 1 day after
explosion (only shown in Figure 3), at the time of maximum light (20
days in the SN's restframe), and at the time $T_b$ the SN's center
reaches minimum impact at $r=b$. Because of the scale in Figures 4 and
6, the supernova's position and size at 1 day don't appear to be
much different than at 20 days and hence are not shown. For both
lensing configurations (left and right columns) row 4 contains 2
panels (left and right) which show the circular photospheres at the
two deflector redshifts, $z_d=0.05$ and $z_d=0.35$ respectively.  The
$z_d=0.05$ panels show the diamond shaped caustics for the two
non-zero shears of rows 1 and 2 and the $z_d=0.35$ panels show the
caustic for the $\gamma\ne 0$ value used in row 3.  If the caustic for
the smaller shear is too small to distinguish from a point at the
origin it is not shown.

We are plotting amplified absolute magnitudes of the supernova in
the V-band, however, due to the redshift of the source, these light
curves would be observed in the I-band. With the source located at
$z_{s}=1$, M$_{V}>-15.5$ is too dim to be seen. Nonetheless, we include
the complete amplification and light curves to show their trends over
a period of 1,000 observer days after the supernova explosion. It should be noticed
that amplification curves are not symmetric (like those of point sources) 
because of the photosphere's expansion.

For many cases, microlensing has a less than dramatic effect on the light curve's shape; 
it simply provides an overall increase in its magnitude, and would be difficult to 
distinguish from amplification due to the galaxy or cluster hosting the lens, see \cite{Saini00}. 
However in more interesting cases, we can easily match the features in the light 
curve to the corresponding features in the amplification curve. 

Figure 3 configurations both show an overall amplification around the supernova's peak brightness as well
as a discernible distortion in the peak itself as a result of a narrow-width rise
in the amplification at this early time (see the inserts in row 1). 
For example, in our model light curve (\S 2.1) the $\Delta M_{20}$
parameter, the decline in magnitudes of the $V$--band light curve
during the first 20 days after maximum light, is 1.10.  This value (in
the SN rest frame) is increased to 1.33, 1.17, and 1.23 for the
respective values of shear $\gamma =0,\ 0.20,$ and 0.40 shown in the
left column of Figure 3, and to 1.57, 1.46, and 1.22 for the
configuration of the right column.  These changes in $\Delta M_{20}$
are rather large.
After the peak phase, shear has little effect
on the remainder of the light curve for either configuration. 

When the photosphere is small compared to the caustic 
structure, high amplification is possible from a neighboring caustic, (\eg \ see the $\gamma=0.2$ caustic
at $z_d=0.05$ of the left panel of the left column or row 4), however,
the intrinsic brightness of the SN is undetectably low then.  
When the SN is near its peak brightness, the photosphere 
has grown to be the size of the caustic structure and the amplification is maximum; 
continued expansion then reduces the net amplification.
 The third row is the result of computing lensed light curves without relative motion and subtracting them from light curves similar to those in the first row, \ie lensed light curves where the relative motion is present. The right column of row 3 shows that relative motion effects can be distinguished from photospheric expansion in some cases, \eg a full magnitude difference can occur at $z_d=0.05$ when relative motion is included.
As expected the magnitude differences diminish with redshift and are practically gone when $z_d=0.35$. The photosphere's 
expansion speed, when projected into the deflector plane, increases with deflector redshift and becomes more important whereas the 
deflector's relative transverse speed would be roughly constant for real lenses (exactly constant for our examples). 
The second and fourth panels of the fourth row show why shear and relative motion become less important with increasing redshift, \ie the photosphere is even larger relative to the caustic structure and motion produces smaller relative displacements. 

The larger mass deflectors of Figures 4 and 5 have larger Einstein ring sizes and hence larger 
caustic dimensions relative to SN displacements and photosphere sizes at any given time. 
For these cases the amplification is larger and shear can change 
the amplification curves significantly; however, Schwarzschild deflectors often produce amplification peaks 
larger than Chang-Refsdal deflectors as is seen in Fig. 4 and in the left column of Fig. 5.

These more massive lenses show a large overall increase in the brightness of the whole
light curve together with occasional second maxima occurring much later than the peak brightness. 
A low shear seems more likely to produce a plateau or a second peak, however, 
the reverse is the case for the configuration shown in the right column of Figure 5. 
For this configuration $v=1.0$ AU/day = 1,730 km/sec is large enough to have the photosphere cross a caustic
while its size remains smaller than the caustic dimension. The second caustic crossing would appear beyond 1000 days.
In the left column where $v=0.5$ AU/day the $\gamma =0.05$ caustic is crossed before 100 days, which alters the SN's peak brightness, but then significant photospheric expansion occurs before reaching the second caustic and results in a long flat amplification curve.
Likewise, for the left column, 2$^{nd}$ row, $\gamma=0.05$ case of  Figure 4 a caustic crossing peak in amplification occurs early and distorts the magnitude peak. 
The source is small when it crosses the caustic after which expansion occurs rapidly causing little more than a rise in the remainder of the light curve. 

Whether or not a second peak occurs in the light curve depends on the 
height and width of the net amplification peak and when it occurs in the life of the SN. 
In general we can say that for small shear the presence of a  peak  in the 
amplification curve occurs around the time of minimun
impact, however, for larger shears, caustic crossings distort the amplification peak
towards the caustic crossing times as seen in Figures 4 and 5. If the caustic crossings coincide with 
minimum impact as in the right column of Fig. 5 when $z_d=0.05$ and $\gamma =0.10$, the total amplification is much enhanced. At lower speeds the
amplification curves flatten out and remove any possibility of a second peak occurring in the light curve.
For example the plateaus in the $\gamma = 0$ curves of row 1 of Figure 4 are still present (but diminished) at $z_d=0.10$ and are gone by $z_d=0.35$. For higher redshift deflectors, relative motion diminishes relative to the 
increased caustic size and photospheric expansion levels the amplification curve.

\section{Conclusion}

We have shown that microlensing can significantly affect light
curves of some cosmologically distant SNe~Ia. We restricted
our calculation to sources at $z_{s}=1$ in the currently accepted
$\Omega _{m}=0.3$, $\Omega _{\Lambda }=0.7$ flat cosmological model.
We found that microlensing can not only increase the magnitude of the
light curve but also can cause a change in its shape. 
Relative transverse motion of the SN and lens, when added to the expanding 
photosphere, 
can result in features such as 
an an enhanced peak brightness with a distorted shape as in Figure 3, 
a post peak plateau as in Figure 4, or even the presence of a wide second peak as in Figure 5.

In the absence of relative lens-source motion \citet{Schneider87}
found distortions to the light curve's peak caused by the supernova's
photosphere expanding into a deflector's critical point as well as into its caustics. 
However, from row 3 of Figures 3-5 it is easy to see
how ignoring the relative motion at small deflector redshifts can
result in underestimating the variety and intensity of lensing effects
on the SNe light curves.  We find that stationary deflectors do not
produce features such as bumps or plateaus as often as the
corresponding moving deflectors. This is due to the fact that the
amplification curves for stationary lenses tend to be flattened and
show less change in amplification as the photosphere expands through
and beyond the deflector.
The effect of amplification by stationary deflectors appears to
produce more of an overall upward shift in the supernova's light
curve. A look at row 3 of Figures 3 through 5 shows that the magnitude
difference of the moving versus stationary cases is measurable for the
deflectors at $z_{d}$=0.05 and 0.10 but not as far as 0.35.  The
magnitude differences fall below 0.1 at $z \sim 0.25$.

Also, it should be pointed out that a supernova lensed by a stationary
deflector is brighter than one lensed by a moving deflector beyond the
time when the projected distance of source and deflector is greater
than the initial value $l_{o}$; see, for instance, the right column of row 3 of Figure 3
where the two $z_d=0.05$ curves go negative at $T=169$ days.

In $\S$ 3 we pointed out, that for microlensing by compact masses
distributed through the cosmos, the optical depth is 0.01
$\left(z_{s}\sim 1\right)$ and might reach 0.1 $\left(z_{s}\sim
2\right)$, implying that the overall chance of a distant SN~Ia being
microlensed is not negligible. Any multi-band supernova survey aimed
at finding supernovae at redshifts around $z=1$ (and above), could
discover and identify \emph{one} microlensed SN Ia event out of
roughly \emph{a hundred} events. However, the low impact parameters
required to produce the special features depicted in $\S$ 4 demand
observation of $\sim 10^{4}$ supernovae at $z_{s}\geqslant 1$.  To see
unusual features such as double peaks, the lensed supernova must be
followed for an extended period of $\sim$ 2 years.  In the Appendix we
have made an optical-depth type estimate to include double peak
events. For microlensing by stars in the bulge of a galaxy at
$z_d=0.05$ we find a max probability of $\sim 1.7\times 10^{-3}$.
This is an ideal deflector distance for observing double peaks due to
transverse motion.  As expected this number is only slightly smaller
than the 27\% optical depth estimate made in $\S 3.3$ for bulge
lensing when corrected for an impact parameter of $u=0.1$. These
estimates can double when the observational bias in favor of amplified
events are taken advantage of (see, for instance, \citet{Gunnar03}).
It is interesting to note that according to OGLE III \citep{Udalski03}
a histogram of $u$ values for Bulge lensing peaks at $u\sim 0.1$,
probably due to amplification biasing.

Notice that we have not taken into account macrolensing  convergence
effects caused by the deflector's host galaxy. The effects of convergence
are relatively easy to include, \eg in our case the amplifications for the Schwarzschild lens given in the second rows of Figure 3-5 
would increase by $\sim$30\% if the lens galaxy had an unexpectedly high transparent 
surface mass density of $\Sigma=$ 1000 M$_{\sun}$pc$^{-2}$ at the image. 

\acknowledgments{}

The authors are pleased to thank Eric Lentz for providing the
limb-darkening curves for model W7 and Zach Blankenship for correcting
an oversight in our work. This work was in part supported by NSF grants
AST-0204771 and AST-0506028, and NASA grants NNG04GD36G NAG5-3505.

\appendix
\section{Appendix}
In this appendix we compute the probability that a source,
followed for a period $T$, impacts a Schwarzschild lens with a reduced
impact parameter less than $u\equiv b/r_E$ and simultaneously moves 
at least a distance $b$ during the period $T$. Such a time 
dependent impact will cause a change in the amplification of 
10\%-50\% depending on the actual impact. The idea here is to estimate
the chance of seeing a distortion in the light curve of a SN whose life 
time is $T_{SN}\sim 200$ days.

We start with a number density $N_d$ of  mass $m$ deflectors (located at a distance $D_d$ 
from the observer) moving with 
relative transverse velocities distributed according to:
\begin{equation}
\frac{dN_d}{dv}=N_d(D_d)\frac{v}{v_{rms}^2}e^{-v^2/2v_{rms}^2}.
\end{equation}
The probability of one of these moving deflectors impacting the line of 
sight to a source at $D_s$  with a reduced 
impact parameter $\leqslant u$ and moving a reduced distance $\geqslant u$ during 
a period $T$ is:
\begin{equation}
\Delta Prob(u,T) =\int^{D_s}_0dD_d\int^{\infty}_{u\,r_E/T}dvN_d(D_d)(\pi u^2r_E^2+2ur_Ev\,T)
\frac{v}{v_{rms}^2}e^{-v^2/2v_{rms}^2}.
\end{equation}
The integrand is the sum over cigar-shaped areas with widths $2ur_E$ and lengths $vT_{SN}+2ur_E$ (traced out by the moving deflectors) . 
The velocity integral can be done easily, 
and if the deflectors are effectively confined to a plane, 
the result can be written as

\begin{equation}
\Delta Prob(u,\xi) = \frac{4G}{c^2}\ \Sigma\,D\, u^2\left\{(\pi+2)
e^{-\xi^2/2}+\sqrt{2\pi}\frac{Erfc(\xi/\sqrt{2})}{\xi}\right\},
\end{equation}
where
\begin{equation}
\xi\equiv \frac{ur_E}{v_{rms}T}= u\,\frac{T_{rms}}{T},
\end{equation}
$\Sigma$ is the projected surface mass density
\begin{equation}
\Sigma \equiv \int^{D_s}_0m\,N_d(D_d)dD_d,
\end{equation}
and $Erfc$ is an error function.

The characteristic crossing time for microlensing is defined by 
$T_{rms}=r_E/v_{rms}=\sqrt{2r_S D}/v_{rms}$ (see $\S$ 3 for definitions)
which for Galaxy bulge-bulge lensing is about 10 days \citep{Udalski03}. 
For a similar 
galaxy at redshift $z=0.05$ 
lensing a distant SN through its bulge, the reduced distance $D$ 
is increased by a factor of $\sim 2.7\times 10^5$ [see $\S$ 3.3 and 
\citet{Han03}] and hence 
$T_{rms}$ increases to $\sim$ 5,200 days. If $u\sim 0.05$ and 
$T=T_{SN}\sim 200$ days, then $\xi \sim$ 1.3
and
$\Delta Prob(0.05,1.3)\sim 1.7\times 10^{-3}$. This particular probability falls 
off by at least an order of magnitude when $u<0.01$ or $u>0.15$.

\clearpage

\figcaption{} Expansion velocity in km~s$^{-1}$ (\emph{left axis}:
filled triangles are observed data and dotted curve is an exponential
fit) and the radius in AU (\emph{right axis}: solid curve assuming
homologous expansion) are plotted against time since explosion.
\label{f1}

\figcaption{} Normalized limb-darkening curves in the U and B bands
calculated for the W7 model at 15 days after explosion, and two curves
$\propto \sqrt{1-r^2/r^2_{SN}}$ for homologous expansion velocities of
13,000 km s$^{-1}$ and 15,000 km s$^{-1}$. 
\label{f2}
 
\figcaption{}
Light curves (\emph{1$^{st}$ row}) and amplification
curves (\emph{2$^{nd}$ row}) of a SN~Ia at $z_{s}=1.0$, microlensed
by a deflector at $z_{d}=0.05$ with a mass of $m_{d}=10^{-3}$ M$_{\odot}$
for two different relative transverse velocities $\mathbf{v}=v\,\hat{\imath}$, impact positions 
 $\mathbf{b}=b\,\hat{\jmath}$, and initial distances $l_0$   (left and right Columns). The un-lensed lightcurve is the solid line in the \emph{1$^{st}$ row}. 
In the \emph{3$^{rd}$ row} the difference in lensed light curves, caused by moving minus non-moving 
deflectors, are shown for $\gamma=0$  at three  redshifts $z_{d}=0.05, 0.10$ and 0.35, and for three shears $\gamma = 0.20, 0.35 (=1.76\times 0.20), 0.63 (=3.15\times 0.20)$
at the same three respective redshifts. Larger $z_d$'s are plotted with thicker lines, $\gamma=0$ as dotted curves and the largest $\gamma$ as dashed curves. 
When a given microlens and its environment is taken from $z=0.05$ 
and placed at 0.10 and 0.35, $\gamma$ scales as  1.76 and 3.15 respectively ($\propto D\equiv D_dD_{ds}/D_s$). 
The \emph{4$^{th}$ row} shows the position and size of the moving supernova's photosphere at three times: 1 day after explosion, at 20
days after the explosion when the SN Ia is at its peak brightness, and at the time $T_b$ its center reaches minimum impact $b$, for two different redshifts $z=0.05$ and $z=0.35$ (left and right panel respectively) for each lensing configuration \ie for each column. The caustics are also shown for the two
shear values 0.20 and 0.40 in the $z=0.05$ panel and for shear value 0.63 in the $z=0.35$ panel
(the linear units are in Einstein ring radii $r_E$).
\label{f3}

\figcaption{}
Similar to Fig. 4, with $m_{d}=1$ M$_{\odot}$, but with $\gamma=$ 0.0, 0.05, and 0.20 for the left column and 
$\gamma=$ 0.0, 0.10, and 0.15 for the right. For the \emph{3$^{rd}$ row} on the right
$\gamma=0.15, 0.26=1.76\times 0.15, 0.47=3.15\times 0.15$.  
In the \emph{4$^{th}$ row} left column the 
caustics are for $\gamma=0.05$ and  0.20 in the left panel, and the same as in Fig. 4 
for the right panel. For the right column shear values 0.10 and 0.15 in the $z=0.05$ panel and for shear value 0.47 in the $z=0.35$ panel. The 1 and 20 day photospheres are not distinguishable at the scale shown. 
\label{f4}

\figcaption{}
Similar to Fig. 5, with $m_{d}=10$ M$_{\odot}$. 
\label{f5}

\clearpage

 \epsscale{.5}
\plotone{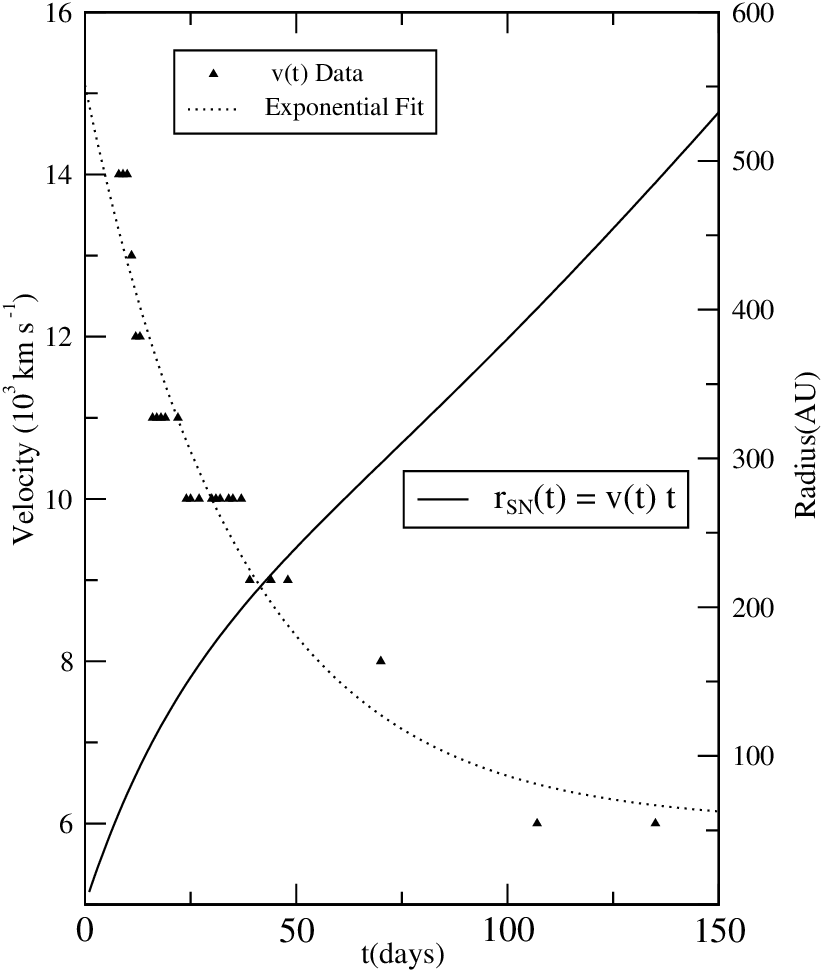}
\begin{center} Figure 1 \end{center}
\vskip .3in
 \epsscale{.6}
\plotone{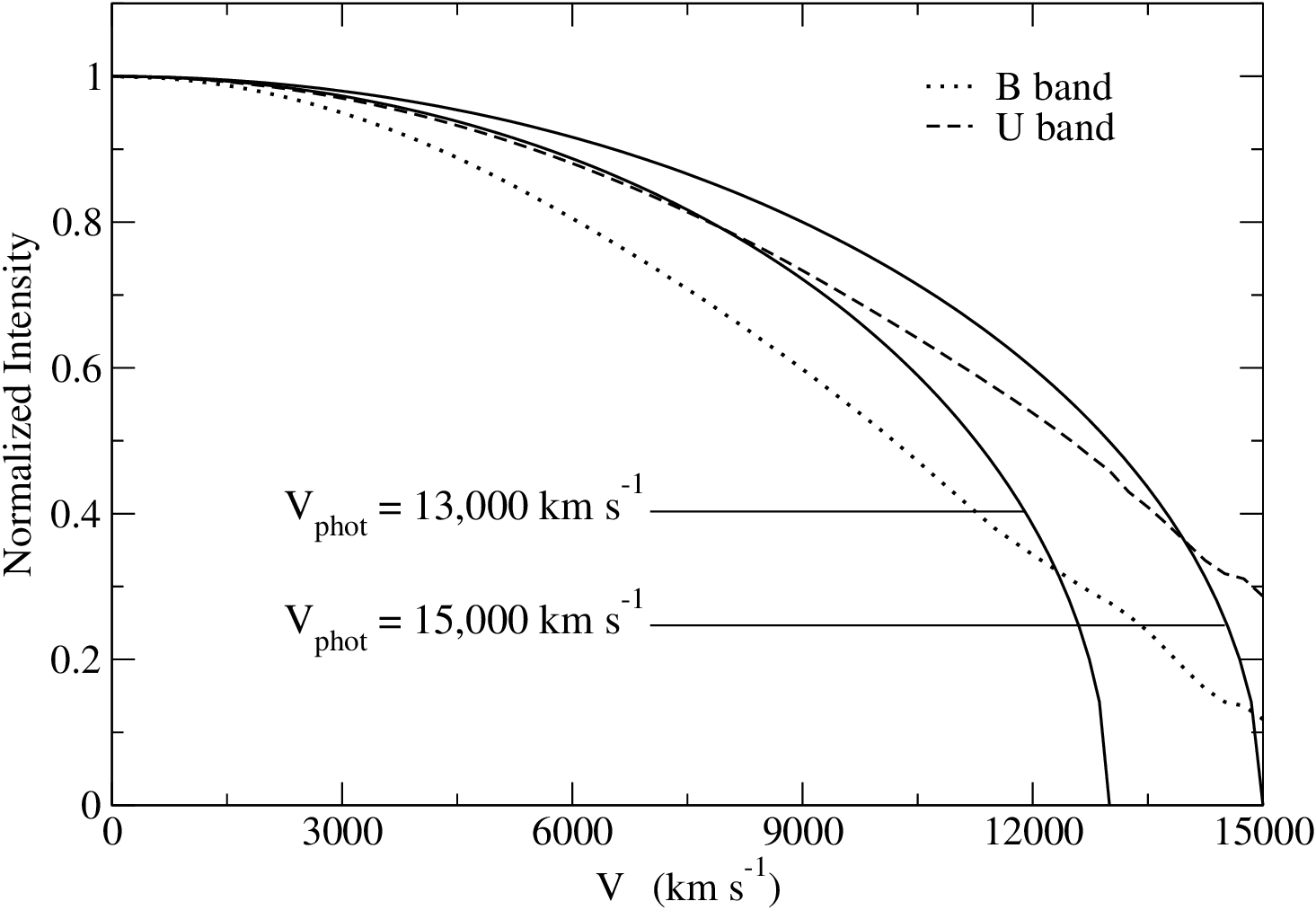}
\begin{center} Figure 2 \end{center}

 \epsscale{1}
\plotone{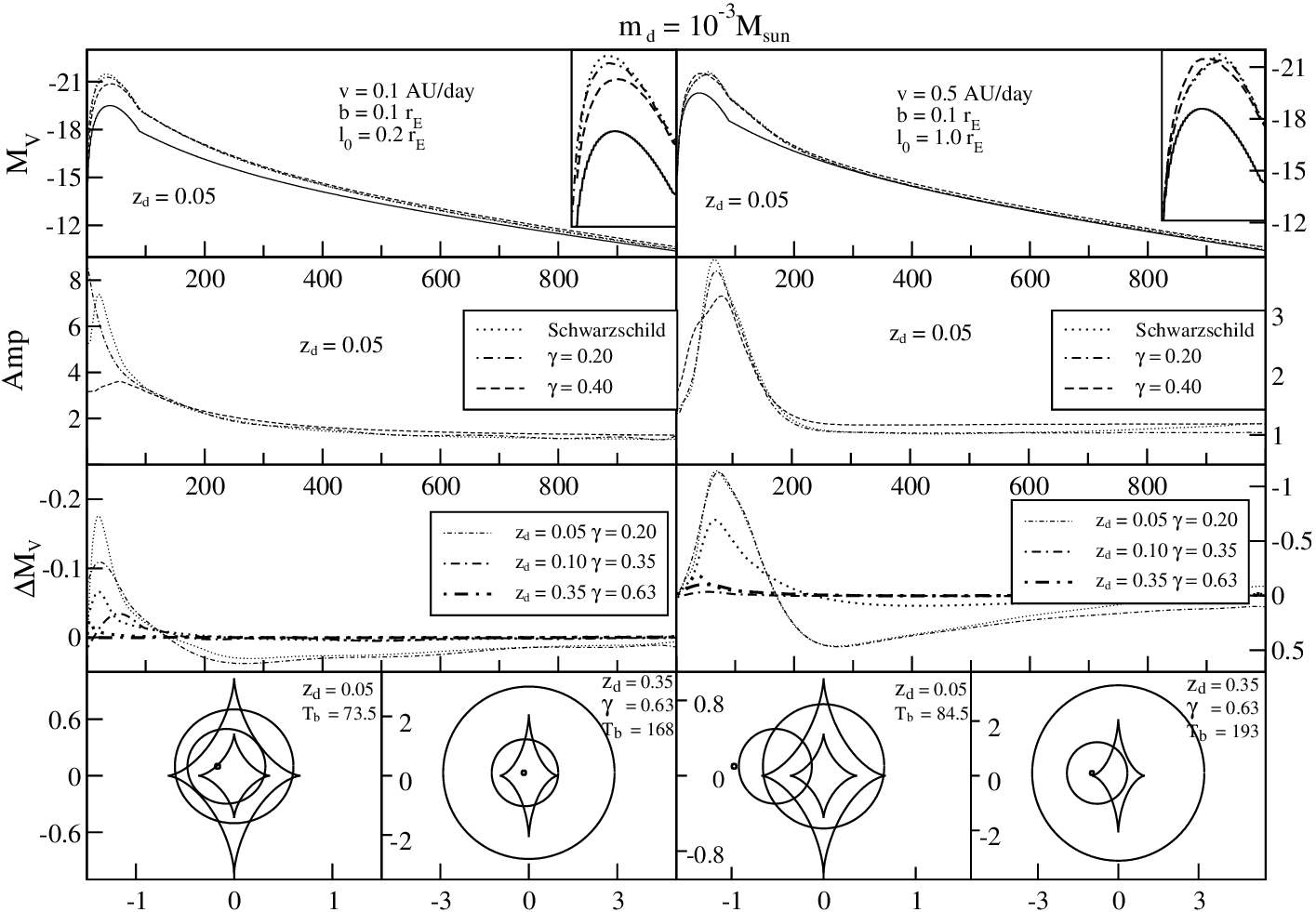}
\begin{center} Figure 3 \end{center}

\plotone{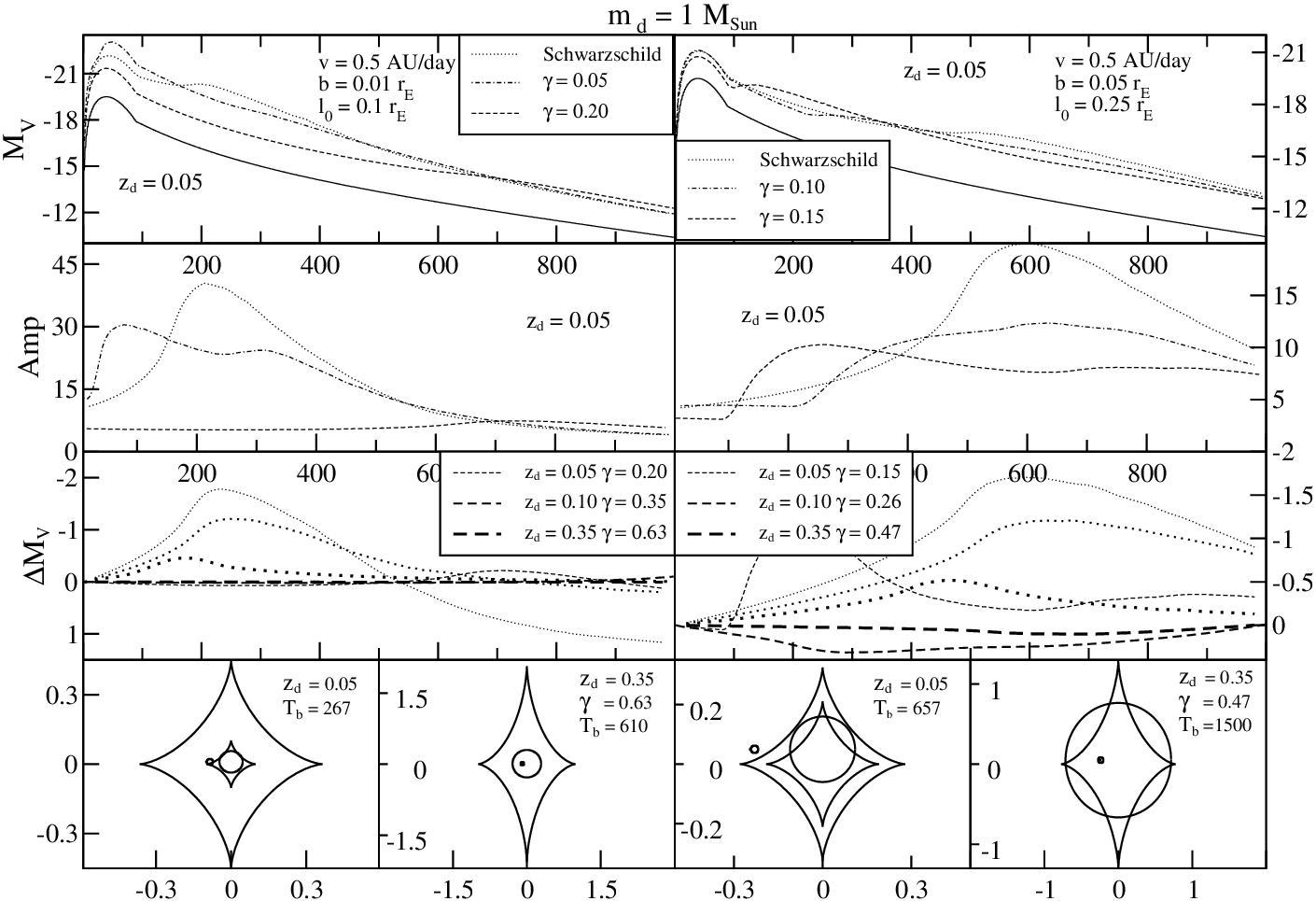}
\begin{center} Figure 4 \end{center}

\plotone{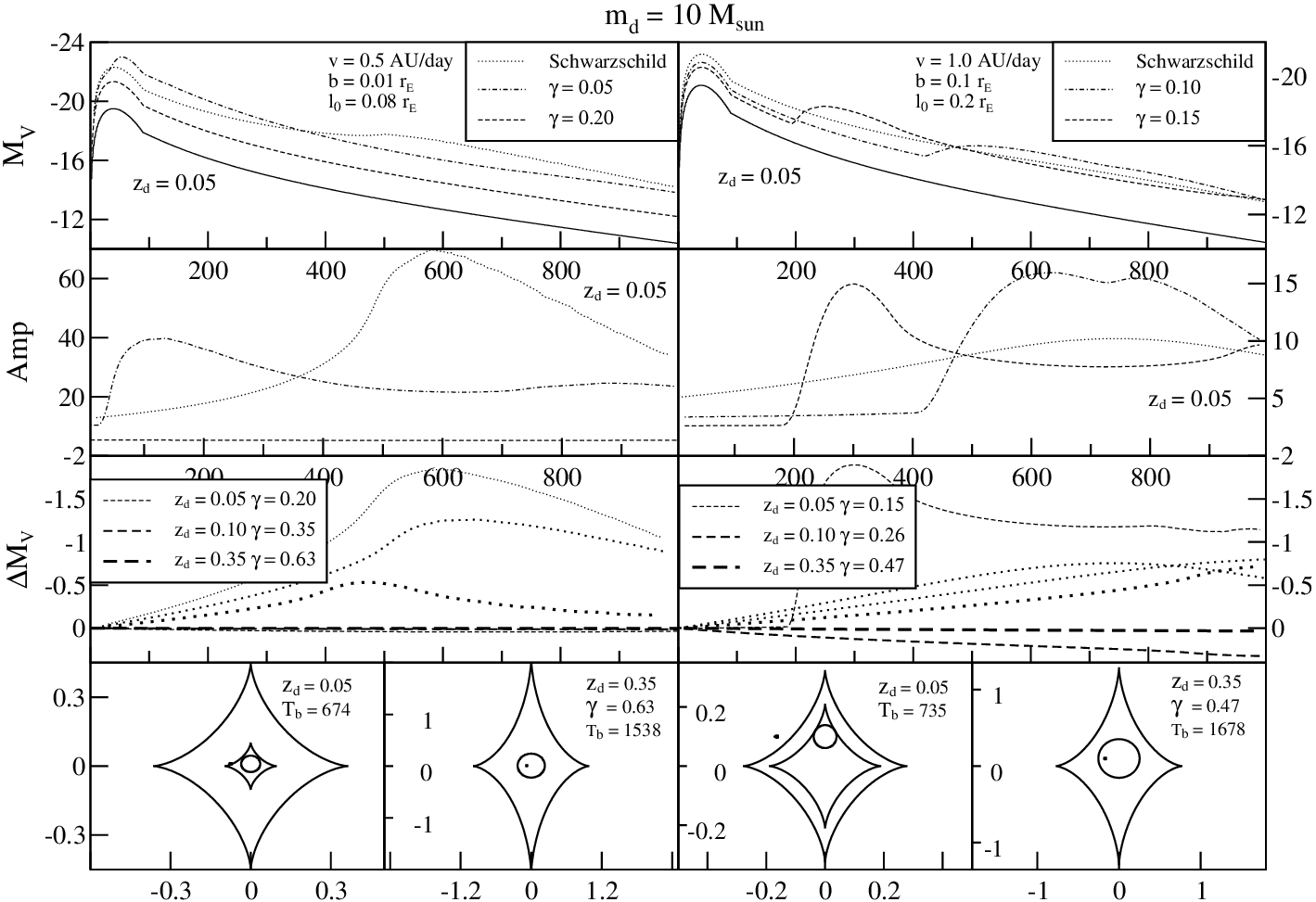}
\begin{center} Figure 5 \end{center}

\end{document}